\documentclass[aps,showpacs,preprintnumbers]{revtex4}
\usepackage{amsmath}
\usepackage{amssymb}
\usepackage{bm}
\usepackage{graphicx}
\usepackage{epstopdf}

\begin{document}

\title{Revisiting fermion helicity flip in Podolsky's Electromagnetism }
\author{Jorge Henrique Sales$^{1}$, A.T. Suzuki,$^{2}$ and Ronaldo Thibes$
^{3}$}
\affiliation{$^{1}${Universidade Estadual de Santa Cruz,\\
Rodovia Jorge Amado km 16} \\ CEP 45662-900 - Ilh\'{e}us, BA,
Brazil.}
\affiliation{$^{2}$Instituto de F\'{\i}sica Te\'{o}rica-UNESP,\\ R. Dr. Bento Teobaldo Ferraz, 271 \\ CEP  01140-070 S\~{a}o Paulo,
SP, Brazil}
\affiliation{$^{3}$Departamento de Ci\^{e}ncias Exatas e Naturais,\\
Universidade Estadual do Sudoeste da Bahia, CEP 45700-000, Itapetinga, BA, Brazil}
\date{\today }

\begin{abstract}
The spin projection of a massive particle onto its direction of motion is called helicity (or ``handedness''). It can therefore be positive or negative. When a particle's helicity changes from positive to negative (or vice-versa) due to its interaction with other particles or fields, we say there is a helicity flip. In this work we show that such helicity flip can be seen for an electron of  $20 MeV$ of energy interacting with a charged scalar meson through the exchange of a virtual photon. This photon {\it does not} necessarily need to be Podolsky's proposed photon; in fact, it is independent of it.
\end{abstract}

\maketitle

\section{Introduction}

As an introduction, we report to the work of Accioly and Mukai \cite{1} previously published back in 1997. In their work they claim that the massive fermions of energy $100 MeV$ do exhibit helicity flip as the particles interact with Podolsky's generalized electromagnetic field.
Their claim is based on and drawn from the three-dimensional diagram plots  shown in figures 7 and 8 in the conclusion section of their article. However, those plots clearly show us that {\em there is no helicity flip} since de graph for their $P$ (fermion's polarization) axis for both diagrams present {\em only} positive values in the range $0 \le P \le 1$. In order to show a helicity filp one has to have in the plot, a graph going from positive values to negative values of $P$.   

In this work we revisit their calculation for the polarization considering a charged fermion being scattered by a charged scalar pion in the Podolsky's electrodynamics \cite{2}. To do that, we consider at the tree level the scattering of a charged fermion by a scalar charged pion, initially at rest in the laboratory frame. For a fermion with energy $100 MeV$ we reproduce their tree-dimensional plot (see Figure \ref{fig:Accioly}). This plot does not evidence helicity flip for the fermion.

However, as we lower the fermion's energy to $20 MeV$ (see Figure \ref{fig:Sales}) we can clearly see the helicity flip for the fermion, from positive values of $P$ going to negative values of $P$ (the bottom right corner of the tree-dimensional graph has values around $P = - 0.5$). 

We analyse the reaction process
\begin{equation}
e^{-}+\pi ^{+}\rightarrow e^{-}+\pi ^{+}  \label{espalhamento}
\end{equation}
where $e^{-}$ represents the electron traveling with energy $E$ towards a pion $\pi ^{+}$ at rest. After the scattering, there emerges an electron in a direction that makes an angle $\theta$ relative to the original direction. This process reveals a change in the polarization of the electron discussed in \cite{1}, using Podolsky's theory \cite{2,6,7,8}. First we obtain the relevant Podolsky's photon propagator which enter in the calculation of the right-handed and left-handed helicities. We show that for the electron with $100 MeV$, there is no helicity flip; however, for lower energy electron with $20 MeV$, helicity flip does really occur. 

\section{Helicity flip}

The covariant Lagrangian density for the theory is given by 
\begin{eqnarray}
{\mathcal L} &=&-\frac{1}{4}F_{\mu\nu}F^{\mu\nu}
+\frac{a^{2}}{2}(\partial _{\alpha }F^{\mu \alpha })^{2}-\frac{1}{2\lambda 
}(\partial _{\mu }A^{\mu })^{2}+(\partial _{\mu}\phi )^{\ast }D^{\mu}\phi -m_{\pi }^{2}\phi ^{\ast }\phi +\overline{\psi }(i\rlap\slash\partial
-m_{e})\psi +e\overline{\psi }\gamma _{\mu }A^{\mu }\psi +  \notag \\
&&+ie(A^{\mu }\phi \partial _{\mu }\phi ^{\ast }-A^{\mu }\phi ^{\ast
}\partial _{\mu }\phi )+e^{2}A_{\mu }\phi ^{\ast }A^{\mu }\phi.
\label{lagrange_total}
\end{eqnarray}

In the Lagrangean density above, we have represented the electron mass by $m_{e}$ and the pion mass by $m_{\pi }$, electrically charged with charges $-e$ and $+e$ respectivelly.  The relevant Feynman rules that we need are
\begin{eqnarray}
V_{A}^{\mu }(p,q) &=&ie\gamma ^{\mu }\text{\:\:is the vertex for the fermions with initial momentum $p$ and final momentum $q$ }  \notag \\
V_{B}^{\mu }(v,w) &=&-ie(v+w)^{\mu }\text{  is the vertex for the pions of initial momentum $v$ and final momentum $w$}  \label{feynman} 
\end{eqnarray}

The propagator for the photon in the Podolsky's theory can be calculated from the Lagrangean density and is given by \cite{1}
\begin{equation}
D_{\mu \nu }(k)=\frac{iM^{2}}{k^{2}(k^{2}-M^{2}+i\varepsilon )}\left[ g_{\mu
\nu }-\frac{1-\lambda (1-\frac{k^{2}}{M^{2}})}{k^{2}}k_{\mu }k_{\nu }\right],
\label{propagador}
\end{equation}
where $k$ is the momentum of the virtual photon exchanged between the electron and the pion and $M^2 \doteq 1/a^2$ is the characteristic Podolsky's parameter.

Without loss of generality, we may consider the effect of negative helicity of an incident electron in a pion at rest  according to the process (\ref{espalhamento}) \cite{1}. The helicity flip for a fermionic particle beam can be calculated by the evaluation of the scattered polarization defined by 
\begin{equation}
P=1-\frac{2N_{\text{right}}}{N_{\text{left}}+N_{\text{right}}},
\label{helicity}
\end{equation}
where $N_{\text{left}}$ is the number of fermions that emerge from the scattering with negative helicity and $N_{\text{right}}$ is the number of scattered fermions with positive helicity.

The polarization as defined by Equation (\ref{helicity}) ranges between $-1\leq P\leq 1$. If there is no positive helicity among the scattered particles, $N_{\rm right} = 0$, and consequently $P=1$ and all the scattered fermions have the same helicity as the incident fermions, that is, negative helicity. We then conclude that there is no helicity flip in this case. On the other hand, if $N_{\rm left} = 0$  (that is, all the scattered fermions have now positive helicity, or opposite helicity as compared to the incident ones)  then $P = -1$ and we conclude that there is a total helicity flip. Evidently, between these two extreme situations, we may find helicity flip occuring partially. The key to determine whether there was or there was not a helicity flip is the behavior of the polarization $P$: a change in the sign of the quantity $P$ as defined in Equation (\ref{helicity}) signals helicity flip; otherwise no helicity flip. 

For the process of scattering (\ref{espalhamento}), we have the analytic expression from the Feynman rules,
\begin{equation*}
N_{\text{right}}=\left| \bar{u}_{\rm R}(q)V_{A}^{\mu }(p,q)u_{\rm L}(p)D_{\mu \nu }(k)V_{B}^{\nu }(v,w)\right|^{2},
\end{equation*}
and
\begin{equation*}
N_{\text{left}}=\left| \bar{u}_{\text{L}}(q)V_{A}^{\mu }(p,q)u_{
\text{L}}(p)D_{\mu \nu }(k)V_{B}^{\nu }(v,w)\right| ^{2},
\end{equation*}
where $D_{\mu \nu }(k)$ is the photon propagator (\ref{propagador}).

Spinors $u_{\text{L}}(p)$ and $u_{\text{R}}(p)$ satisfy the property (we use Feynman slash notation for contraction with Dirac gamma matrices)
\begin{equation}
u_{\text{L,R}}(p)=\left( \frac{1+\gamma ^{5}\,\rlap\slash S_{\text{L,R}}}{2}
\right) u_{\text{L,R}}(p),  \label{prop_1}
\end{equation}
and the polarization vectors are given by
\begin{equation}
S_{\text{L,R}}^{\mu }(p)=\left( \frac{\mp \left| \vec{p}\,
\right| }{m_{e}},\frac{\mp p_{0}}{m_{e}}\frac{\vec{p}}{
\left\vert \vec{ p}\,\right\vert }\right).  \label{polarizacao}
\end{equation}

We thus have
\begin{equation}
N_{\text{Right}}=\left\vert \frac{ieM^2}{k^2(k^2 - M^2)}
\bar{u}_{\text{R}}(q)\gamma ^{\mu }u_{\text{L}}(p)\left\{ g_{\mu \nu }-
\frac{1-\lambda (1-\frac{k^{2}}{M^{2}})}{k^{2}}k_{\mu }k_{\nu }\right\}
(v+w)^{\nu }\right\vert ^{2}.  \label{nr}
\end{equation}

Using now the following properties for fermions with momentum $p$ and $q$, we have:
\begin{equation*}
\left. u_{\text{L,R}}(p)\right\vert _{\alpha }\left. \bar{u}_{\text{
L,R}}(p)\right\vert _{\beta }=\left\{ \left( \frac{\rlap\slash p+m_{e}}{2m_{e}}
\right) \left( \frac{1+\gamma _{5}\rlap\slash S_{\text{L,R}}(p)}{2}\right).
\right\} _{\alpha \beta }
\end{equation*}

The $\alpha $ and $\beta $ are matrix indices and we may rewrite the expression as
\begin{eqnarray*}
\label{Nright}
N_{\text{Right}} &=&\frac{e^{2}M^4}{(k^2)^2(k^2-M^2)^2}Tr\left\{ \left( \frac{\rlap\slash 
p+m_{e}}{2m_{e}}\right) \left( \frac{1+\gamma _{5}\rlap\slash S_{\text{R}}(q)
}{2}\right) \rlap\slash Q\times \right. \\
&&\left. \left( \frac{\rlap\slash p+m_{e}}{2m_{e}}\right) \left( \frac{%
1+\gamma _{5}\rlap\slash S_{L}(p)}{2}\right) \rlap\slash Q\right\},
\end{eqnarray*}
where $Q=v+w$. Using the trace properties for gamma matrices, we have
\begin{eqnarray}
N_{\text{Right}} &=&{\mathbb P}_{\text{El}}\left\{ \frac{-1}{m_{e}^{2}}t_{\rm lab}\left[ 
\sqrt{E_{\rm lab}^{2}-m_{e}^{2}}(E+2m_{\pi })-E_{\rm lab}\sqrt{E^{2}-m_{e}^{2}}\cos
\theta \right] \right. \times  \notag \\
&&\times \left[ \sqrt{E^{2}-m_{e}^{2}}(-E_{\rm lab}+2m_{\pi })+E_{\rm lab}\sqrt{%
E^{2}-m_{e}^{2}}\cos \theta \right] +\frac{\left( 4m_{\pi
}^{2}-t_{\rm lab}\right) }{m_{e}^{2}}  \notag \\
&&\left( E\sqrt{E_{\rm lab}^{2}-m_{e}^{2}}-E_{\rm lab}\sqrt{E^{2}-m_{e}^{2}}\cos
\theta \right) \left( E_{\rm lab}\sqrt{E^{2}-m_{e}^{2}}-E\sqrt{%
E_{\rm lab}^{2}-m_{e}^{2}}\cos \theta \right) +  \notag \\
&&+\frac{1}{2}\left[ \left( 4m_{\pi }^{2}-t_{\rm lab}\right) t_{\rm lab}-\left(
s_{\rm lab}-u_{\rm lab}\right) ^{2}\right] \left( 1+\frac{\sqrt{E^{2}-m_{e}^{2}}%
\sqrt{E_{\rm lab}^{2}-m_{e}^{2}}(EE_{\rm lab}\cos \theta )}{m_{e}^{2}}\right) +
\label{nr_final} \\
&&-\frac{1}{m_{e}^{2}}\left( s_{\rm lab}-u_{\rm lab}\right) \left[ \left( \sqrt{%
E_{\rm lab}^{2}-m_{e}^{2}}(E+2m_{\pi })-E_{\rm lab}\sqrt{E^{2}-m_{e}^{2}}\cos \theta
\right) \right. \times  \notag \\
&&\left( E_{\rm lab}\sqrt{E^{2}-m_{e}^{2}}-E\sqrt{E_{\rm lab}^{2}-m_{e}^{2}}\cos
\theta \right) +  \notag \\
&&+\left. \left. \left( E\sqrt{E_{\rm lab}^{2}-m_{e}^{2}}-E_{\rm lab}\sqrt{%
E^{2}-m_{e}^{2}}\cos \theta \right) \left[ \sqrt{E^{2}-m_{e}^{2}}%
(-E_{\rm lab}+2m_{\pi })+E\sqrt{E_{\rm lab}^{2}-m_{e}^{2}}\cos \theta \right] \right]
\right\}.  \notag
\end{eqnarray}

In Equation (\ref{Nright}), for the sake of convenience,  we have introduced the shorthand notation  for the overall coefficient as $${\mathbb P}_{\rm El}\doteq \frac{e^2M^4}{4t^2(t-M^2)^2m_e^2}.$$

For $N_{\text{left}}$ we have similarly
\begin{eqnarray}
N_{\text{left}} &=&{\mathbb P}_{\text{El}}\left\{ \left( 4m_{\pi }^{2}-t_{\rm lab}\right)
t_{\rm lab}-\left( s_{\rm lab}-u_{\rm lab}\right) ^{2}+\frac{1}{m_{e}^{2}}t_{\rm lab}\left[ 
\sqrt{E_{\rm lab}^{2}-m_{e}^{2}}(E+2m_{\pi })-E_{\rm lab}\sqrt{E^{2}-m_{e}^{2}}\cos
\theta \right] \right. \times  \notag \\
&&\left[ \sqrt{E^{2}-m_{e}^{2}}(-E_{\rm lab}+2m_{\pi })+E_{\rm lab}\sqrt{%
E^{2}-m_{e}^{2}}\cos \theta \right] +  \notag \\
&&-\frac{\left( 4m_{\pi }^{2}-t_{\rm lab}\right) }{m_{e}^{2}}\left( E\sqrt{%
E_{\rm lab}^{2}-m_{e}^{2}}-E_{\rm lab}\sqrt{E^{2}-m_{e}^{2}}\cos \theta \right)
\times  \notag \\
&&\left( E_{\rm lab}\sqrt{E^{2}-m_{e}^{2}}-E\sqrt{E_{\rm lab}^{2}-m_{e}^{2}}\cos
\theta \right) -\frac{1}{2}\left[ \left( 4m_{\pi }^{2}-t_{\rm lab}\right)
t_{\rm lab}-\left( s_{\rm lab}-u_{\rm lab}\right) ^{2}\right] \times  \notag \\
&&\left( 1+\frac{\sqrt{E^{2}-m_{e}^{2}}\sqrt{E_{\rm lab}^{2}-m_{e}^{2}}%
(EE_{\rm lab}\cos \theta )}{m_{e}^{2}}\right) -\frac{1}{m_{e}^{2}}\left(
s_{\rm lab}-u_{\rm lab}\right) \times  \label{nl_final} \\
&&\left[ \left( \sqrt{E_{\rm lab}^{2}-m_{e}^{2}}(E+2m_{\pi })-E_{\rm lab}\sqrt{%
E^{2}-m_{e}^{2}}\cos \theta \right) \left( E_{\rm lab}\sqrt{E^{2}-m_{e}^{2}}-E%
\sqrt{E_{\rm lab}^{2}-m_{e}^{2}}\cos \theta \right) \right. +  \notag \\
&&-\left. \left. \left( E\sqrt{E_{\rm lab}^{2}-m_{e}^{2}}-E_{\rm lab}\sqrt{%
E^{2}-m_{e}^{2}}\cos \theta \right) \left[ \sqrt{E^{2}-m_{e}^{2}}%
(-E_{\rm lab}+2m_{\pi })+E\sqrt{E_{\rm lab}^{2}-m_{e}^{2}}\cos \theta \right] \right]
\right\}.  \notag
\end{eqnarray}

In the above expressions, we have used the Mandelstam variables in the laboratory frame
\begin{eqnarray*}
s_{\rm lab} &=&m_{e}^{2}+m_{\pi }^{2}+2Em_{\pi }, \\
t_{\rm lab} &=&2m_{\pi }\left( E_{\rm lab}-E\right) , \\
u_{\rm lab} &=&2\left( m_{e}^{2}+m_{\pi }^{2}\right) -s_{\rm lab}-t_{lab,}
\end{eqnarray*}%
where
\begin{equation}
E_{\rm lab}=\frac{\left( m_{\pi }E+m_{e}^{2}\right) \left( E+m_{\pi }\right)
+\cos \theta \left( E^{2}-m_{e}^{2}\right) \sqrt{m_{\pi }^{2}-m_{e}^{2}\sin
^{2}\theta }}{\left( E+m_{\pi }\right) ^{2}+\left( E^{2}-m_{e}^{2}\right)
\cos ^{2}\theta }  \label{energia_lab}
\end{equation}%
and $E$ is the fermion energy. We note that the scattered fermion polarizarion $P$ as given in Equation (\ref{helicity}) does not depend on the overall coefficient ${\mathbb P}_{\text{El}}$ that appears in front of the $N_{\rm right}$ and $N_{\rm left}$. This means that the polarization does not depend on the characteristic Podolsky's parameter $M^2 \doteq 1/a^2$. This means our result is independent of the character of the virtual photon propagator, whether it is given by the usual Maxwell's theory or by the Podolsky's approach to electromagnetism.

\begin{figure}[!htb]
\centering
\includegraphics[scale=.7]{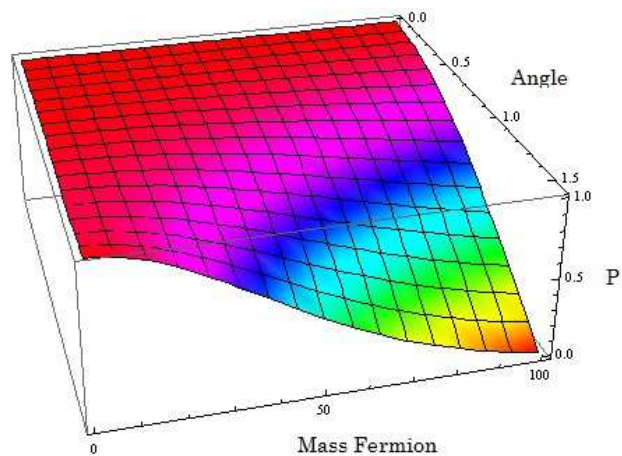}
\caption{Electron scattering with initial total energy equal to 100 MeV.}
\label{fig:Accioly}
\end{figure}

We are working in the reference frame where the pion is at rest, that is,  $w^{\mu }=(m_{\pi
},0,0,0)$, with initial energy for the electron $E=100$ Mev. In this case, we reproduce exactly the result obtained in reference \cite{1}, and our graph in Figure \ref{fig:Accioly} is precisely their Figure 7. Neither their Figure 7 nor Figure 8 of their article show that for fermion mass of the order of electron mass, there is a helicity flip as they claimed. Even when the fermion mass gets larger than the electron mass there is only a change in the direction of the helicity (not exactly paralel to the momentum direction), but still the helicity does not undergo sign change, so there is no flip.

\begin{figure}[!htb]
\centering
\includegraphics[scale=.7]{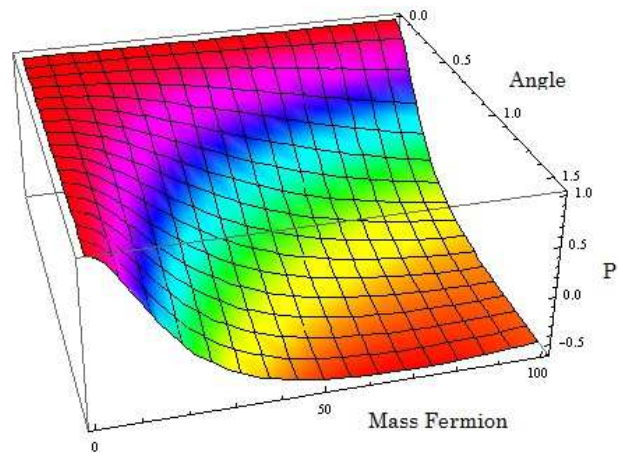}
\caption{Electron scattering with initial total energy equal to 20 MeV.}
\label{fig:Sales}
\end{figure}

In our graph of Figure \ref{fig:Sales}, however, we have a curious effect: If we lower the electron energy from $100$ MeV to $20$ MeV, we observe a range of negative polarization, implying that helicity flip is occuring for increasing mass of the fermion.


\section{Conclusion}

We have shown that the authors of reference \cite{1} have missed their point as far as their claim of helicity flip for electron interacting with Podolsky's photon is concerned.Their affirmation of helicity flip for the electron undergoing interaction with a scalar charged meson with an exchange of Podolsky's type photon cannot be substantiated. Their graph does not corroborate for such an assertion and neither their allegation that this effect is due to the Podolsky's photon is warranted. 

However, we have shown that for a particular lower energy for the electron, i.e., 20 MeV instead of 100 MeV considered in reference \cite{1} we clearly see a helicity flip. Morever, this flip has nothing to do with the interacting intermediate photon being {\em a la} Podolsky or not, for the polarization factor does not depend on any Podolsky's factor $a^2$. 

\vspace{.7cm}

\textbf{Acknowledgments:} ATS thanks the hospitality of the Physics Department of  NCSU, Raleigh, NC and gratefully acknowledges a research grant from FAPESP, BPE process 2014/20829-2. J.H. Sales wishes to thank NCSU and IFT for the hospitality while this work was in progress and acknowledges research grants from FAPESB-Propp-00220.1300.1088 and Capes-FAPESB proc. 3336/2014. RT thanks the hospitality of NCSU and partial financial support from CNPq, process 202141/2015-2.

\end{document}